\documentclass[twocolumn,epjc3]{svjour3}
\usepackage[T1]{fontenc}
\usepackage[utf8]{inputenc}
\usepackage[british]{babel}
\usepackage{amsmath,bbm,braket,graphicx}
\usepackage{slashed}
\declareslashed{}{/}{0.10}{0}{p}
\declareslashed{}{/}{0.08}{0}{\hat{p}}
\declareslashed{}{/}{0.12}{0}{\hat{n}}
\newcommand\UnitOp{\mathbbm{1}}                      
\DeclareMathOperator\Tr{Tr}                          
\DeclareMathOperator\CC{C}                           
\newcommand\IPhantLo{_{\vphantom{I}}}                
\newcommand\IPhantHi{^{\vphantom{I}}}                
\usepackage[dvips]{hyperref}
\hypersetup{%
  pdfauthor={Philip G. Ratcliffe <philip.ratcliffe@uninsubria.it>},
  pdftitle={Single Transverse-Spin Asymmetries in Drell-Yan Processes},
  pdfsubject={submitted to the European Physical Journal, section C},
  pdfkeywords={transverse spin, parton model, twist three},
}
\usepackage[dvips]{feynmp}
\usepackage[nofmf]{feynmphax}
\journalname{Eur. Phys. J. C}
\begin{document}
\title{Single Transverse-Spin Asymmetries in Drell--Yan Processes}
\author{%
  Gabriele Re Calegari\thanksref{addr1}
\and
  Philip G. Ratcliffe\thanksref{email1,addr1,addr2}
}
\thankstext{email1}{e-mail: philip.ratcliffe@uninsubria.it}
\institute{%
  Dipartimento di Scienza e Alta Tecnologia,
  Universit\`{a} degli Studi dell'Insubria---sede di Como,
  via Valleggio~11, 22100 Como, Italy
  \label{addr1}
\and
  Istituto Nazionale di Fisica Nucleare---sezione di Milano Bicocca,
  piazza della Scienza 3, 20126 Milano, Italy
  \label{addr2}
}
\date{Received: date / Accepted: date}
\maketitle
\abstract{%
  The asymmetry in the angular distribution of Drell--Yan dilepton pairs in collisions where just one nucleon is transversely polarised has been examined in the literature with a variety of results, differing mainly by factors of two. We re-evaluate the asymmetry via \mbox{twist-3} contributions in collinear factorisation. In order to allow complete and in-depth comparison with existing calculations, we supply all calculational details.
}
\begin{fmffile}{fmffiles/fmf}
\section{Introduction}

In this paper we focus our attention on the study of the cross-section for Drell--Yan \cite{Drell-Yan} processes in which an unpolarised proton and a transversely polarised antiproton annihilate to produce a dilepton pair.
In particular, we shall calculate the single-spin asymmetry (SSA) defined as
\begin{equation}
  A_N =
  \frac{\displaystyle
         \frac{d\sigma( \vec{S}_T)}{d\Omega dQ^2}
       - \frac{d\sigma(-\vec{S}_T)}{d\Omega dQ^2}}
       {\displaystyle
         \frac{d\sigma( \vec{S}_T)}{d\Omega dQ^2}
       + \frac{d\sigma(-\vec{S}_T)}{d\Omega dQ^2}},
\end{equation}
where $\vec{S}_T$ is the polarisation of the antiproton in the plane orthogonal to its motion. The study of this asymmetry is of particular interest for a variety of reasons. Firstly, a number of experiments propose to measure it. Secondly, given the small number of quarks involved in such processes, SSAs are ideal observables to further test factorisation mechanisms in pQCD. Lastly, SSAs are related to particular parton distribution functions that describe the inner structure of hadrons.

The motivation to review this specific topic is the diversity of the results found in literature. The first calculation was performed in \cite{Hammon}, where the authors obtained the following result:
\begin{multline}
  A_N =
  -\frac{1}{Q} \, \frac{\sin2\theta \sin\phi_S}{1 + \cos^2\theta}
\\
  \times
  \frac{\sum_q e^2_q \int dx \,
        \left[ \displaystyle T_F^q(x,x) - x\frac{d}{dx}T_F^q(x,x) \right]
        f^{\bar{q}}(x')}
       {\sum_q e^2_q \int dx \, f^q(x) f^{\bar{q}}(x')},
\label{eq:Hammon}
\end{multline}
where $f^q$ is the twist-two (unpolarised) distribution for a quark of flavour $q$, the momentum fraction of the antiquark is defined by $x'=Q^2/(xs)$, $s=(P+P')^2$ is the CM frame energy squared and $Q^2$ the momentum squared of the virtual photon. The correlator $T_F(x,x)$ is a twist-three quark--quark--gluon spin-dependent matrix element (given \emph{e.g.} in \cite{Hammon}), which will be described in detail later.

The derivative term in the numerator of \eqref{eq:Hammon} was later doubted and the result given in \cite{Boer1,Boer2} is
\begin{equation}
  A_N =
  -\frac{1}{Q} \frac{\sin2\theta \sin\phi_S}{1 + \cos^2\theta}
  \frac{\sum_q e^2_q \int dx \, T_F^q(x,x) f^{\bar{q}}(x')}
       {\sum_q e^2_q \int dx \, f^q(x)     f^{\bar{q}}(x')}.
\label{eq:result1}
\end{equation}
The question of the derivative was further addressed in \cite{Teryaev:2000pi} without clear conclusions while in \cite{Boer3} its absence was argued owing to the absence of so-called double soft-gluon poles in this calculation.

Result \eqref{eq:result1} was obtained by expanding the hadronic tensor in terms of \mbox{twist-3} correlators depending on the hadron transverse spin. A parallel approach may be found in \cite{Ma}, in which factorisation is in terms of transverse momentum dependent quark--quark correlators. The final result is, however, in contrast with the previous two results.

A further attempt was made in \cite{Teryaev}, in which result \eqref{eq:result1} was found to be multiplied by a factor two. The SSA was recently recalculated in \cite{Metz}, where the authors noted that transverse-momentum dependence must be taken into account not only in the hadronic tensor, but also in the leptonic tensor; indeed, $\vec{p}_T$ flows from the hadronic to the leptonic part via the virtual photon. The result found there was half of \eqref{eq:result1}.
In \cite{Ma:2012ph} the calculation was performed using partonic states and the result obtained coincides with that of \cite{Metz}. The motivation given there for the difference with respect to the earlier papers appears dissimilar in form although the interplay between the transverse-momentum dependence in both the hadronic and leptonic tensors is central.

In order to attempt to clarify the situation, we have repeated the calculation taking into account the observations of \cite{Metz}; we obtain, however, a result which is two times \eqref{eq:result1}. In this paper we shall define all the partonic distribution functions and quantities used and show all calculational details.

The rest of the paper is organised as follows. In the next section we provide the definitions of the kinematical variables used and we define the quark--quark and quark--quark--gluon correlators with respect to the associated parton distributions. In section~3 we describe the Collins--Soper (CS) reference frame \cite{Collins-Soper}, which we use in the calculation, and the transverse-spin vector in these coordinates. We then focus our attention on the question of gauge invariance of the correlators and suppression of gauge-link operators. In the final two sections we examine the collinear expansion and show all calculational details involved in reaching the final result.

\section{Notation and Definitions}

In this section we review the notation and definitions used in this paper. Let us start by noting that in the ultrarelativistic regime the momentum carried by a quark is essentially in the hadron direction, or rather, in the beam direction. Therefore, given a pair of vectors:
\begin{equation}
  \hat{p}^\mu = \tfrac{1}{\sqrt{2}}(1,0,0,1)
  \ \text{and} \
  \hat{n}^\mu = \tfrac{1}{\sqrt{2}}(1,0,0,-1),
\end{equation}
taking the third component along the beam direction, we may define
\begin{equation}
  p^\mu = \Lambda \, \hat{p}^\mu
  \ \text{and} \
  n^\mu = \Lambda^{-1} \, \hat{n}^\mu,
\end{equation}
and their light-cone projections
\begin{equation}
\begin{alignedat}{2}
  p^+ &= \tfrac{1}{\sqrt{2}} (p^0+p^3) = \Lambda     , \\
  p^- &= \tfrac{1}{\sqrt{2}} (p^0-p^3) = 0           , \\
  n^+ &= \tfrac{1}{\sqrt{2}} (n^0+n^3) = 0           , \\
  n^- &= \tfrac{1}{\sqrt{2}} (n^0-n^3) = \Lambda^{-1}.
\end{alignedat}
\end{equation}
The specific value of $\Lambda$ will determine the chosen hadron reference frame.

Consequently, a generic vector may be written as $P^\mu=(p^+,p^-,\vec{P}_T)$, with $\vec{P}_T=(p_1,p_2)$ while a generic scalar product is $P{\cdot}P'=P^+P'^-+P^-P'^+-\vec{P}_T{\cdot}\vec{P}'_T$. The power of this notation lies in the fact that, by choosing the beam direction as $p^+$, as in a light-cone coordinate system, this component will dominate in scalar products and we may easily isolate effects induced by transverse spin and momentum. With these definitions, we may write the hadron momentum as
\begin{equation}
  P^\mu = p^\mu + \tfrac{1}{2} M^2 \, n^\mu,
\label{eq:P}
\end{equation}
This relation is true not only in the hadron rest frame ($\Lambda=M/\sqrt{2}$, with $M$ the hadron mass), but also in the infinite-momentum frame ($\Lambda=P^+\rightarrow\infty$).

In the second part of this section we define one of the fundamental objects used in the study of hadronic scattering processes: the quark--quark correlator. Let us start by noting that the transition matrices that describe the passage from an hadronic state with definite momentum and spin $\ket{P,S}$ to a generic state $\ket{X}$ are of the form $\bra{X}\psi_i\ket{P,S}$. As a consequence, to obtain a transition probability, we need to multiply by the complex conjugate and sum over all intermediate states. It is therefore natural to define the quark--quark correlator
\begin{equation}
  \phi_{ij}(P,S,p) =
  \int \frac{d^4z}{(2\pi)^4} \, e^{ip{\cdot}z}
  \Bra{P,S} \bar{\psi_i}(0) \psi_j(z) \Ket{P,S}.
\label{eq:correlator}
\end{equation}
Correlators written in this form are not measurable in physical processes and our interest will therefore be to connect this object with observables such as the hadron momentum and spin.

Firstly, we note that the correlator is a $4{\times}4$ Dirac matrix, a Lorentz scalar and depends on three independent physical quantities $[P,S,p]$, where $P$ is the hadron momentum, $S$ its spin and $p$ the quark momentum. In parallel, there are only three independent matrices, they are $[\UnitOp,\gamma^\mu,\gamma_5]$. Let us now multiply this set of vectors and matrices by each other respecting the following properties:
\begin{itemize}
\item
  Invariance under parity. Parity changes the signs of momenta, but not of axial vectors such as spin. As a consequence, helicity, defined as $\lambda=\frac{\vec{P}{\cdot}\vec{S}}{|\vec{P}|}$, flips sign under such transformations. For parity to be conserved, the following must be valid:
  \begin{equation}
    \phi(P,S,p) =
    \gamma^0 \, \phi(\tilde{P},-\tilde{S},\tilde{p}) \, \gamma^0.
  \end{equation}
\item
  Invariance under charge conjugation and time reversal. This condition implies that:
  \begin{equation}
    \phi^{*}(P,S,p) =
    \gamma_5 \CC \phi(\tilde{P},\tilde{S},\tilde{p}) \CC^\dag \gamma_5,
  \end{equation}
  where $\CC=i\gamma^0\gamma^2$ is the charge-conjugation operator. This is obvious since the transformation does not alter the spatial part of the spin operator and helicity does not change sign.
\item
  The correlator must be Hermitian, which implies:
  \begin{equation}
    \phi^\dag(P,S,p) = \gamma^0 \phi(P,S,p) \gamma^0.
  \end{equation}
\item
  The expansion must be linear in spin. This condition arises from the fact that Lorentz invariance implies that the hadronic tensor must be linear in the spinors $u(P,S)$ and $\bar{u}(P,S)$. As a consequence, the tensors constructed with this requirement are spin independent or linear in spin:
\begin{subequations}
\begin{align}
  \bar{u}(P,S) \gamma^\mu u(P,S) &= 2P^\mu
\\[-2ex]
\intertext{or}
\notag\\[-5ex]
  \bar{u}(P,S) \gamma^\mu\gamma_5 u(P,S) &= 2M S^\mu \quad (S^2=-1).
\end{align}
\end{subequations}
\end{itemize}

The set of combinations of vectors based on these requirements is as follows:
\begin{equation}
  \lbrace
    \slashed{p}, \, \slashed{P}, \, \slashed{S}\gamma_5, \, p{\cdot}S\gamma_5
  \rbrace.
\end{equation}
Note that we need not consider all possible products between $p$, $P$ and $S$, but only those linearly dependent on the spin $S$. This is because the coefficients $A_i$ multiplying each structure in the complete expansion will depend on $p{\cdot}P$ and $p^2$.

Let us now construct the most general expansion of the correlator over the basis obtained by multiplying the elements of the set of vectors and matrices. We must pay attention to the condition mentioned earlier. The maximum number of products is limited by the fact that the product of a quantity with itself does not generate any new Dirac structure. In fact, we have $\UnitOp^2=\UnitOp$, $\slashed{p}^2=p^2\UnitOp$, $\slashed{P}^2=P^2\UnitOp$, $(\slashed{S}\gamma_5)^2=\UnitOp$ while $(p{\cdot}S\gamma_5)^2$ is forbidden by the request of linearity in spin. As a consequence, if $r$ is the number of basic elements multiplying each other, we obtain
\begin{subequations}
\begin{align}
  r=1: & \quad
  \UnitOp, \; \slashed{p}, \; \slashed{P}, \; \slashed{S}\gamma_5, \;
  p{\cdot}S\gamma_5;
\\
  r=2: & \quad
  \slashed{p}\slashed{P}, \; \slashed{p}\slashed{S}\gamma_5, \;
  p{\cdot}S\slashed{p}\gamma_5, \; \slashed{P}\slashed{S}\gamma_5, \;
  p{\cdot}S\slashed{P}\gamma_5;
\\
  r=3: & \quad
  \slashed{p}\slashed{P}\slashed{S}\gamma_5, \;
  p{\cdot}S\slashed{p}\slashed{P}\gamma_5.
\end{align}
\end{subequations}
The method produces $12$ independent structures; not all, however, fulfill the requirements of time-reversal invariance and hermiticity. To see which satisfy these conditions it is convenient to find out which of these terms include a double or triple slash. To do this we use the relations
\begin{subequations}
\begin{align}
  \slashed{a} \slashed{b} &= -i\sigma^{\mu\nu} a_\mu b_\nu + a{\cdot}b
\\
\intertext{and}
  \gamma^5 \slashed{a} \slashed{b} \slashed{c} &=
  i\epsilon^{\sigma\mu\nu\rho} \gamma_\sigma a_\mu b_\nu c_\rho
\notag
\\
  & \hspace*{5em}
  + a{\cdot}b \, \gamma^5 \slashed{c} - b{\cdot}c \, \gamma^5 \slashed{a}
  + c{\cdot}a \, \gamma^5 \slashed{b},
\end{align}
\end{subequations}
in which the only interesting terms are the first as the others have already been taken into account (recall that $a{\cdot}b$ is proportional to $\UnitOp$).

We now have the correlator expansion:
\begin{multline}
  \phi(P,S,p) =
  A_1 M \UnitOp + A_2 \slashed{P} + A_3 \slashed{p}
  + A_6 M \slashed{S} \gamma_5
\\
  \hspace*{4em}
  + \frac{A_7}{M} p{\cdot}S \, \slashed{P} \gamma_5
  + \frac{A_8}{M} p{\cdot}S \, \slashed{p}\gamma_5
  + i A_9 \sigma^{\mu\nu} \gamma_5 S_\mu P_\nu
\\
  + i A_{10} \sigma^{\mu\nu} \gamma_5 S_\mu p_\nu
  + i \frac{A_{11}}{M^2} \sigma^{\mu\nu} \gamma_5 \, p{\cdot}S \, p_\mu P_\nu.
  \label{eq:sviluppogenerale}
\end{multline}
In parallel, the correlator is a matrix in Dirac space; it  may therefore be expanded over an orthonormal basis of $\gamma$ matrices to obtain:
\begin{multline}
  \phi(P,S,p) =
  \tfrac{1}{2}
  \Big[
    S\,\UnitOp + iP \gamma^5 + V_\mu \gamma^\mu
\\
    + A_\mu \gamma^5 \gamma^\mu + \tfrac{i}{2} T_{\mu\nu} \sigma^{\mu\nu}
  \Big],
  \label{eq:corr}
\end{multline}
where the letters $S$, $P$, $V$, $A$ and $T$ indicate the types of currents: scalar, pseudoscalar, vector, axial and tensor respectively. Comparing expressions \eqref{eq:sviluppogenerale} and \eqref{eq:corr}, we obtain
\begin{subequations}
\begin{align}
  S &= \tfrac{1}{2} \Tr\!\big( \phi \big) \equiv A_1 M,
\\
  V_\mu &=
  \tfrac{1}{2} \Tr\!\big( \gamma_\mu \phi \big) \equiv A_2 P_\mu + A_3 p_\mu,
\\
  A_\mu &= \tfrac{1}{2} \Tr\!\big( \gamma_\mu \gamma_5 \phi \big)
\notag
\\
  &\equiv A_6 M S_\mu + \frac{A_7}{M} \, p{\cdot}S \, P_\mu
  + \frac{A_8}{M} \, p{\cdot}S \, p_\mu,
\\
  T_{\mu\nu} &= -\tfrac{1}{2} i \Tr\!\big( \sigma_{\mu\nu} \gamma_5 \phi \big)
\notag
\\
  &\equiv i A_9 P_{[\mu}S_{\nu]} + i A_{10} p_{[\mu}S_{\nu]}
    + i \frac{A_{11}}{M^2} \, p{\cdot}S \, P_{[\mu} p_{\nu]},
\end{align}
\end{subequations}
where $[\,\dots]$ around indices indicates antisymmetrisation, \emph{i.e.} $P_{[\mu}S_{\nu]}\equiv{}P_\mu{}S_\nu-P_\nu{}S_\mu$.

We may simplify these expressions, noting in this case that $P^\mu\sim{}P^+$, $p^\mu=xP^\mu$, $MS^\mu=\lambda{}P^\mu+MS^\mu_T$ and we may neglect terms in $M^2/P^+$.
As a consequence, neglecting quark transverse momentum, we can rewrite these equations in terms of the proton momentum and transverse spin only. Let us define new coefficients $D_i$ depending only on the quark momentum fraction $x$. Keeping only the dominant terms, we redefine:
\begin{subequations}
\begin{align}
  V_\mu &= D_1 P_\mu, \\
  A_\mu &= \lambda D_2 P_\mu, \\
  T_{\mu\nu} &= D_3 P_{[\mu}S_{T\nu]}.
\end{align}
\end{subequations}
As a consequence, it is possible to rewrite the correlator in terms of the structure functions $f(x)$, $\Delta{}f(x)$ and $\Delta_Tf(x)$. We obtain:
\begin{align}
  \phi(x) &\equiv
  \int \frac{d^4p}{(2\pi)^4} \,
  \delta\!\left(\frac{p}{P^+} - x\right) \phi(P,S,p)
\notag
\\
  &
  = \tfrac{1}{2}
  \Big[
    f(x) \slashed{P} + \Delta f(x) \lambda \gamma^5 \slashed{P}
    + \Delta_T f(x) \gamma^5 \slashed{S}_T \slashed{P}
  \Big].
  \label{eq:corrsviluppo}
\end{align}

Thus far we have worked at the twist-two level. Alternatively, following \cite{Kogut}, it is possible to project the Dirac spinors onto `good' and `bad' components and show that they are connected by gluonic fields whenever the quark involved in the process is off-shell. In particular, defining the Hermitian projectors $\mathbbm{P}_{\pm}=\frac{1}{2}[1\pm\gamma^3\gamma^0]$, the `good' components are $\psi_+=\mathbbm{P}_+\psi$ and the `bad' are $\psi_-=\mathbbm{P}_-\psi$. The following relation is also true:
\begin{align}
  \psi_- &=
  \frac{1}{4i} \int d\xi \, \epsilon(x^3-\xi)
  \Big[ (i\partial_j - eA_j) \gamma^j + m \Big] \gamma^0 \psi_+
\notag
\\
  &\equiv
  \frac{1}{4i} \int d\xi \, \epsilon(x^3-\xi) \slashed{D}_T \gamma^0 \psi_+.
\label{eq:relpsipiumeno}
\end{align}
We observe that the relation between `$+$' and `$-$' components depends on a term containing the covariant derivative, and thus on the Dirac equation. Therefore, if the quark is on-shell, it is possible to rewrite $\psi_-$ in terms of $\psi_+$ and the quark mass, but if the quark is off-shell there are no relations between `good' and `bad' components and the addition of new degrees of freedom that are not eliminable by a different choice of gauge is necessary. These are given by the interaction of the transverse gluon field with the quark.

In order to isolate these dependences and create general \mbox{twist-3} distributions, it is necessary to insert the $n$ components into the Sudakov decomposition made earlier. Using relation \eqref{eq:P} together with the expansion
\begin{equation}
  S_\mu = S{\cdot}n \, p_\mu + S{\cdot}p \, n_\mu + S_{T\,\mu}
\end{equation}
in \eqref{eq:corrsviluppo}, with convenient redefinitions, we obtain the following set of \mbox{twist-2} to 4 distribution functions \cite{Jaffe-Ji}:\footnote{\relax
  Note that in \cite{Jaffe-Ji} the normalisation $S^2=-M^2$ is adopted.
}
\begin{subequations}
\begin{multline}
  \frac{1}{2} \int \frac{d\lambda}{2\pi} \, e^{i\lambda x}
  \Bra{P,S} \bar\psi(0) \gamma_\mu \psi(\lambda n) \Ket{P,S}
\\
  \equiv f(x) \, p_\mu + f_4(x) M^2 n_\mu,
\end{multline}
\begin{multline}
  \frac{1}{2} \int \frac{d\lambda}{2\pi} \, e^{i\lambda x}
  \Bra{P,S} \bar\psi(0) \gamma_\mu \gamma_5 \psi(\lambda n)\Ket{P,S}
\\
  \equiv g_1(x) M S{\cdot}n \, p_\mu + g_T(x) M S_{T\mu}
       + g_3(x) M^3 S{\cdot}n \, n_\mu,
\end{multline}
\begin{multline}
  \frac{1}{2} \int \frac{d\lambda}{2\pi} \, e^{i\lambda x}
  \Bra{P,S} \bar\psi(0) i\sigma_{\mu\nu} \gamma_5 \psi(\lambda n) \Ket{P,S}
\\
  \equiv h_1(x) [S_{T\mu} p_\nu - S_{T\nu} p_\mu]
  + h_L(x) M^2 [p_\mu n_\nu - p_\nu n_\mu] S{\cdot}n
\\
  + h_3(x) M^2 [S_{T\mu} n_\nu - S_{T\nu} n_\mu].
\end{multline}
\end{subequations}

\begin{figure}
  \centering
  \input{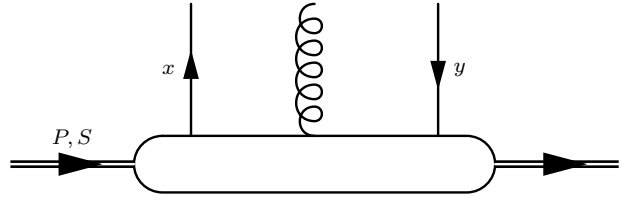}
  \caption{The Born diagram for quark--quark--gluon correlation matrix.}
  \label{fig:qqg-amplitude}
\end{figure}

Wishing now to also display the dependence of the \mbox{twist-3} correlators on the gluonic fields, we expand over independent distribution functions. Let us start by redefining the quark--quark--gluon correlation matrix
\begin{multline}
  \phi_{Dij}^\alpha(x,y) =
  \int \frac{d\lambda}{2\pi} \, \frac{d\eta}{2\pi} \, e^{i\lambda x + i\eta(y-x)}
\\
  \times
  \Bra{P,S} \bar{\psi}_j(0) i D^\alpha(\eta n) \psi_i(\lambda n) \Ket{P,S}
\end{multline}
In general this matrix will be associated with the Born diagram shown in
Fig.~\ref{fig:qqg-amplitude}.

In this case $x$ and $y$ represent the momentum fractions carried by the quarks. More precisely, $x$ is the momentum fraction carried by the quark on the left of the diagram and $y$ on the right. Consequently, $x-y$ is the momentum fraction carried by the gluon. From now on an integral is implied over the quark and gluon momenta with the restrictions $\delta(xP-p)$ and $\delta(yP-p')$, where $p$ and $p'$ are the quark momenta.

We observe that this matrix contains a covariant derivative and thus separates into two parts. The first contains the operator $\bar\psi\partial_\alpha\psi$ and the second the gluonic field $\bar\psi{}A_\alpha\psi$. As already seen for the twist-two distribution, in order to obtain a twist-$\tau$ distribution it is only necessary to consider $\tau$ partons with dynamically independent polarisations. If this independence does not hold, new higher-order contributions will be generated. In order to respect this independence therefore, we project out the ``+'' components from such a correlation matrix and, for the gauge choice $A^+=0$, only transverse gluon polarisations survive.

To perform this projection, we insert $\slashed{n}$ and restrict the index $\alpha$ to be transverse. Consequently, we may define the distributions $G(x,y)$ and $\tilde{G}(x,y)$ for the vector and axial-vector projections of $\phi_{Dij}^\alpha$:
\begin{subequations}
\begin{multline}
  \frac{1}{2} \int \frac{d\lambda}{2\pi} \, \frac{d\eta}{2\pi} \,
  e^{i\lambda x + i\eta(y-x)}
\\
  \hspace*{4em}
  \times
  \Bra{P,S}
    \bar\psi(0) i D_T^\alpha(\eta n) \slashed{n} \psi(\lambda n)
  \Ket{P,S}
\\
  \equiv
  i\epsilon_T^{\alpha\beta} S_\beta \, G(x,y)
  + \dots
\label{eq:G}
\end{multline}
and
\begin{multline}
  \frac{1}{2}\int \frac{d\lambda}{2\pi} \, \frac{d\eta}{2\pi} \,
  e^{i\lambda x + i\eta(y-x)}
\\
  \hspace*{4em}
  \times
  \Bra{P,S}
    \bar\psi(0) i D_T^\alpha(\eta n) \slashed{n} \gamma_5 \psi(\lambda n)
  \Ket{P,S}
\\
  \equiv S_T^\alpha \, \tilde{G}(x,y) + \dots,
\label{eq:G1}
\end{multline}
where $\epsilon_T^{\alpha\beta}\equiv\epsilon^{+-\alpha\beta}$ and the dots denote higher-twist contributions. Note with our normalisation $G(x,y)$ and $\tilde{G}(x,y)$ have dimensions of a mass.

The origin of these two possible choices of expansion over momentum and spin vectors lies in the possible gluon polarisation degrees of freedom. In fact, in the first definition the gluon polarisation is orthogonal to the hadron spin and momentum, in the second it is aligned with the spin. Demanding hermiticity of the operators, we find that $G(x,y)$ is antisymmetric under the interchange $x\leftrightarrow{}y$ whereas $\tilde{G}(x,y)$ is symmetric. Note moreover that invariance under time reversal requires both to be real. Analogously, considering also the tensor interaction, we may define:
\begin{multline}
  \frac{1}{2}\int \frac{d\lambda}{2\pi} \, \frac{d\eta}{2\pi} \,
  e^{i\lambda x + i\eta(y-x)}
\\
  \hspace*{4em} \times
  \Bra{P,S}
    \bar\psi(0)
      \sigma^{\mu\nu} \gamma_5 i D_T^\alpha(\eta n) \slashed{n}
    \psi(\lambda n)
  \Ket{P,S}
\\[1ex]
  \hspace*{2em} \equiv
  \big[ g^{\mu\alpha} p^\nu + g^{\nu\alpha} p^\mu \big] S{\cdot}n \, H(x,y)
\\
  + i\epsilon^{\mu\nu\alpha\beta} E(x,y) \, p_\beta/M + \dots
\label{eq:H}
\end{multline}
\end{subequations}

The relations between definitions (\ref{eq:G}--c) and \eqref{eq:corrsviluppo} have been studied extensively in the literature, in particular in \cite{Boer1}. We have the following:
\begin{subequations}
\begin{align}
  M g_T(x) &=
  \frac{1}{2x} \int dy \,
  \big[ \tilde{G}(x,y) + \tilde{G}(y,x)
\notag
\\
  & \hspace*{6em}
  + G(x,y) - G(y,x) \big],
\\
  M h_L(x) &= \frac{1}{x} \int dy \, \big[ H(x,y) + H(y,x) \big]
\\[-1.5ex]
\intertext{and}
\notag\\[-6.5ex]
  M e(x) &= \frac{1}{x} \int dy \, \big[ E(x,y) - E(y,x) \big].
\end{align}
\end{subequations}

\section{Reference frame and transverse momentum}

In this section we focus our attention on the choice of the reference frame. As the polarisation is defined in the LAB frame transversely to the beam direction, it is natural to choose the Collins--Soper reference frame. With this choice we can isolate transverse-momentum effects, not only in the hadronic tensor, but also in the leptonic tensor. This frame is defined by the following:
\begin{equation}
\begin{alignedat}{2}
  T^\mu &\equiv \frac{q^\mu}{\sqrt{Q^2}},
\\
  Z^\mu &\equiv \frac{2}{\sqrt{Q^2+Q^2_T}}
  \left[ q_{P_2} \tilde{P}_1^\mu - q_{P_1} \tilde{P}_2^\mu \right],
\\
  X^\mu &\equiv
  -\frac{Q}{Q_T} \frac{2}{\sqrt{Q^2+Q^2_T}}
  \left[ q_{P_2} \tilde{P}_1^\mu + q_{P_1} \tilde{P}_2^\mu \right],
\\[1ex]
  Y^\mu &\equiv \epsilon^{\mu\nu\rho\sigma} T_\nu Z_\rho X_\sigma,
\end{alignedat}
\end{equation}
where
\begin{subequations}
\begin{equation}
  \tilde{P_{1,2}^\mu} \equiv \frac{1}{\sqrt{s}}
  \left[ P_{1,2}^\mu - \frac{q{\cdot}P_{1,2}}{q^2} q^\mu \right]
  \equiv \frac{1}{\sqrt{s}} \Big[ P_{1,2}^\mu - \xi q^\mu \Big]
\end{equation}
and
\begin{equation}
  q_{P_{1,2}} = \frac{q{\cdot}P_{1,2}}{\sqrt{s}}.
\end{equation}
\end{subequations}

\begin{figure}
  \centering
  \includegraphics[width=0.45\textwidth]{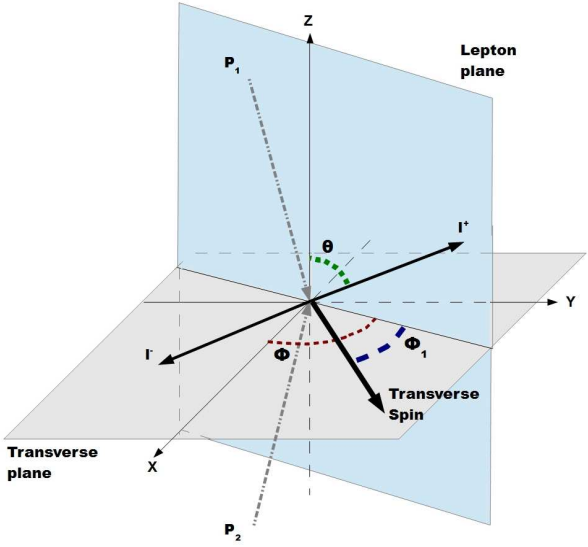}
  \caption{A schematic representation of the Collins--Soper reference frame.}
  \label{fig:CS-frame}
\end{figure}

Let us give a geometric interpretation of this frame: $\xi$ represents the photon momentum fraction in the beam direction. Thus, subtracting $\xi{}q^\mu$ from the proton momentum, we find all the dependence on the transverse momentum of the struck quark. As for the basis vectors, $Z$ indicates the direction of the beam in the absence of transverse momentum, $X$ the transverse-momentum direction, $T$ the photon direction and $Y$, again transverse, completes the set of orthonormal vectors. A simplified representation of this frame is displayed in Fig.~\ref{fig:CS-frame}.

In this frame we can also define the angles we shall use in the calculation of the SSA. In particular, $\theta$ represents the angle between the $Z$ axis and the direction of the outgoing leptons, $\phi_1$ is the angle between the lepton plane and the transverse-spin direction, and $\phi$ is the angle between the lepton plane and the $X$ axis.

As we have seen, the polarisation of the hadron is defined in the LAB frame, we thus now wish to show how the spin vector transforms on moving from the LAB to the CS frame. Let us start by noting that this transformation is possible with two Lorentz boosts in succession. The first is in the beam direction in order to set $Q_3$ to zero, the second is in the $Q_T$ direction to cancel this component too and leave only the temporal component~$Q$. After this transformation it is clear that $Q^2$ corresponds to the mass of the dilepton pair. We represent the spin vector before the boost as:
\begin{align}
  S_T^\mu &= (0,S_x,S_y,0)
\notag
\\
  &= (0,|S|\cos(\phi-\phi_S),|S|\sin(\phi-\phi_S),0).
\end{align}
Applying the two boosts we obtain
\begin{equation}
  S_T^{\mu\text{CS}} =
  \left(
    \begin{array}{c}
      \frac{|Q_T|}{Q} |S|\cos(\phi - \phi_S) \\[1ex]
      \frac{\sqrt{Q^2 + Q_T^2}}{Q} |S|\cos(\phi - \phi_S) \\[1ex]
      |S|\sin(\phi - \phi_S) \\[1ex]
      0
    \end{array}
  \right),
\end{equation}
where we may however neglect terms in $Q_T^2$. We thus see that in this frame the spin vector acquires a temporal component. However, it is easy to show that, after contraction with the leptonic tensor, this has no physical consequence.

In the final part of this section we illustrate the expansion of the leptonic tensor over the CS basis vectors. Since $T$ lies along the photon direction, each of the two leptons carries away half the photon momentum. The canonical decomposition is therefore valid:
\begin{multline}
  l_{\pm}^\mu = \tfrac{1}{2}q^\mu \pm \tfrac{1}{2} Q
  \Big[
    \sin\theta \cos\phi \hat{X}^\mu
\\
  + \sin\theta \sin\phi \hat{Y}^\mu + \cos\theta \hat{Z}^\mu
  \Big],
\end{multline}
Using $L^{\mu\nu}=l_+^{\lbrace\mu}l_-^{\nu\rbrace}-\tfrac{1}{2}Q^2g^{\mu\nu}$ (with $l_+^{\lbrace\mu}l_-^{\nu\rbrace}\equiv l_+^\mu{}l_-^{\mathstrut\nu}+l_-^\mu{}l_+^{\mathstrut\nu}$) and defining $T^\mu\equiv{}q^\mu/\sqrt{Q^2}$, we obtain
\begin{multline}
  L^{\mu\nu} = \frac{Q^2}{2} \Big[ T^\mu T^\nu - g^{\mu\nu} \Big]
\\
  - \frac{Q^2}{2}
  \Big[
    \sin^2\theta \cos^2\phi \hat{X}^\mu\hat{X}^\nu
  + \sin^2\theta \sin^2\phi \hat{Y}^\mu\hat{Y}^\nu
\\
  \hspace*{3.5em}
  + \cos^2\theta \hat{Z}^\mu\hat{Z}^\nu
  + \tfrac{1}{2}\sin^2\theta \sin2\phi
    \hat{X}^{\lbrace\mu} \hat{Y}^{\nu\rbrace}
\\
  + \tfrac{1}{2}\sin2\theta \cos\phi \hat{X}^{\lbrace\mu}\hat{Z}^{\nu\rbrace}
  + \tfrac{1}{2}\sin2\theta \sin\phi \hat{Y}^{\lbrace\mu}\hat{Z}^{\nu\rbrace}
  \Big].
\end{multline}
We note that this tensor is totally symmetric.

\section{Gauge invariance in twist-three correlators}

We now analyse the question of gauge invariance. The gauge choice $A^+=0$, is made for two reasons: firstly, examining expression \eqref{eq:correlator}, we note that $z$ is a space--time variable. Therefore, in order to connect the two different space--time points, we must insert a gauge-link operator into the correlator in the following way
\begin{equation}
  \Bra{P,S} \bar{\psi}_i(0) \, L[0,z] \, \psi_j(z) \Ket{P,S},
\end{equation}
where the gauge link is
\begin{equation}
  L[0,z] = \exp\!\left[ -ig \int_0^z d\eta \, \eta{\cdot}A(\eta) \right].
\end{equation}
The choice $A^+=0$ sets this operator to $\UnitOp$, as the dominant direction of $z$ is the ``+'' direction.

Secondly, in this gauge it is simple to transform a correlator of the form
\begin{multline}
  \phi^\alpha_{A \,ij}(P,S,p)
\\
  = \int \frac{d^4p}{(2\pi)^4} \, e^{ip{\cdot}z}
  \Bra{P,S} \bar{\psi_i}(0) g A^\alpha \psi_j(z) \Ket{P,S}
\label{eq:corrA}
\end{multline}
into a gauge-invariant expression. This may be achieved by noting that
\begin{equation}
  F^{+\alpha} = \partial^+ A_T^\alpha
\end{equation}
and thus, after integration by parts, we arrive at
\begin{equation}
  (x-y) \, \phi_A^\alpha(x,y) = -i\phi_F^\alpha(x,y),
\label{eq:AtoF}
\end{equation}
where $\phi^\alpha_F$ is obtained by replacing $A^\alpha$ with $F^{+\alpha}$ in Eq.~\eqref{eq:corrA}. Using this simple rule, we may then make the substitutions $G_A(x,y)\rightarrow{}G_F(x,y)$ and $\tilde{G}_A(x,y)\rightarrow\tilde{G}_F(x,y)$, where $G_F$ \emph{etc.} are defined by the same replacements in \eqref{eq:G} and \eqref{eq:G1}.

Finally, in order to compare clearly with our calculation, we note that the matrix element $T_F$ appearing in the asymmetry reads
\begin{multline}
  T_F(x,y) =
  \int \frac{d\lambda}{2\pi} \frac{d\eta}{2\pi} e^{i\lambda x + i \eta (y-x)}
\\
  \times
  \bra{PS}
    \bar\psi(0)
    \gamma ^+ \epsilon_T^{\mu\nu} S_{T\nu} gF^+_{\;~\mu}(\eta n)
    \psi(\lambda n)
  \ket{PS}.
\label{eq:T_F}
\end{multline}
Thus, from the structure of Eqs.~\eqref{eq:G}, \eqref{eq:AtoF} and \eqref{eq:T_F}, we see that the function $G_F$ is identical to $T_F$, having the same dependence on transverse spin, with the substitution $A^\alpha\to{}F^{+\alpha}$.

\section{General form of the cross-section}

We begin this section by giving the standard expression for the Drell--Yan cross-section:
\begin{equation}
  \frac{d\sigma}{d^4q d\Omega} =
  \frac{\alpha^2_\text{EM}}{s\,Q^4} \, L_{\mu\nu} \, W^{\mu\nu}.
\label{eq:sezurto}
\end{equation}
The factorisation theorem provides the possibility to divide the hard-scattering part from the soft part. Thus, defining $\phi_1$ and $\phi_2$, the correlators for the proton and the antiproton, we may write the hadronic tensor in the following general form:
\begin{multline}
  W^{\mu\nu} = \frac{1}{N_c} \sum_a e_a^2
  \int \frac{d^4p_1}{(2\pi)^4} \, \frac{d^4p_2}{(2\pi)^4} \,
  \delta^4(p_1+p_2-q)
\\
  \times
  \Tr\!\big( \phi_1 \gamma^\mu \bar\phi_2 \gamma^\nu \big).
\end{multline}
For the case in which we consider an extra exchanged gluon, we must include the gluon propagator in the trace, as we shall in the next section. Note that this general form contains $\delta^4(p_1+p_2-q)$, which includes both the longitudinal and transverse components.

We now perform the collinear expansion of the cross-section. To do this, we expand in a neighbourhood of momentum transverse to the direction of the hadron motion. Calling this momentum $p_T$, we note that in the CS frame its only non-zero components are along the $X$ and $Y$ directions and it is thus of the form $\alpha\hat{X}^\sigma+\beta\hat{Y}^\sigma$. Understanding the derivative as a gradient, we have:
\begin{align}
  L_{\mu\nu} W^{\mu\nu}
  &= L_{\mu\nu} W^{\mu\nu} |_{\vec{p}_T=0}
\notag
\\
  &\hspace*{1em}
  - \frac{Q^2}{2} \frac{\sin2\theta}{2}
  \frac{\partial}{\partial\vec{p}_T}
  \left[\vphantom{\frac00} \cos\phi Z_{\{\mu} X_{\nu\}} W^{\mu\nu} \right.
\notag
\\
  &\hspace*{8em}
  +
  \left.\vphantom{\frac00} \sin\phi  Z_{\{\mu} Y_{\nu\}} W^{\mu\nu}
  \right]_{\vec{p}_T=0} \hspace*{-2.2em} \cdot \vec{p}_T
\notag
\\
  &= L_{\mu\nu} W^{\mu\nu} |_{\vec{p}_T=0}
\notag
\\
  &\hspace*{1em}
  - \frac{Q^2}{2} \frac{\sin2\theta}{2}
  \left[\vphantom{\frac00}
    \cos\phi Z_{\{\mu} X_{\nu\}} + \sin\phi  Z_{\{\mu} Y_{\nu\}}
  \right]
\notag
\\
  &\hspace*{5em}
  \times
  \left[
    W^{\mu\nu} +
    \left(\frac{\partial W^{\mu\nu}}{\partial\vec{p}_T}\right)_{\!\!\vec{p}_T=0}
    \hspace*{-1.9em} \cdot \vec{p}_T
  \right].
\label{eq:ex}
\end{align}

We therefore now need to differentiate the hadronic tensor with respect to the transverse momentum:
\begin{multline}
  \frac{\partial W^{\mu\nu}}{\partial\vec{p}_T} =
  \int \frac{d^4p_1}{(2\pi)^4} \, \frac{d^4p_2}{(2\pi)^4} \,
\\
  \times
  \frac{\partial}{\partial\vec{p}_T}
  \Big(
    \delta^4(p_1+p_2-q) \,
    \Tr\!\big( \phi_1 \gamma^\mu \phi_2 \gamma^\nu \big)
  \Big),
\end{multline}
which, in light-cone coordinates, is equivalent to
\begin{multline}
  \frac{\partial W^{\mu\nu}}{\partial\vec{p}_T} =
  \int \frac{dp_1^+ d\vec{p}^2_{1T}}{(2\pi)^4} \,
       \frac{dp_2^- d\vec{p}^2_{2T}}{(2\pi)^4} \,
\\
  \hspace*{5em}
  \times
  \bigg[
    \frac{\partial\delta^2(q_T)}{\partial\vec{p}_T} \,
      \Tr\!\big( \phi_1 \gamma^\mu \bar\phi_2 \gamma^\nu \big)
\\
    + \delta^2(q_T) \, \frac{\partial}{\partial\vec{p}_T}
      \Tr\!\big( \phi_1 \gamma^\mu \bar\phi_2 \gamma^\nu \big)
  \bigg].
\end{multline}
Integrating the first member by parts gives the second with the opposite sign and so they cancel exactly. Thus, no linear contributions in $\vec{p}_T$ to the DY process survive and therefore the derivative found in \cite{Hammon} is absent.

\section{Calculation of the single-spin asymmetry}

\begin{figure}
  \centering
  \input{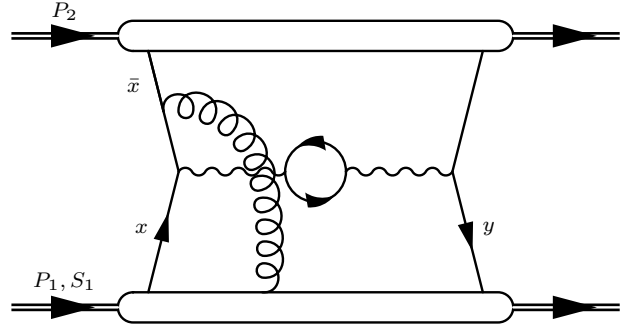}
  \caption{
    The \mbox{twist-3} contribution to the Drell--Yan process (to which must also be added the Hermitian conjugate).
  }
  \label{fig:DY-twist3}
\end{figure}

In this section we show all details of our calculation. We start by noting that, via the optical theorem, we need an imaginary part to obtain a non-zero SSA. Such a contribution can only come from diagrams that involve a single gluon exchange between the polarised hadron and the quark coming from the other hadron. The diagram relevant here is shown in Fig.~\ref{fig:DY-twist3}.
We now write an analytic expression for the hadronic tensor using the factorisation theorem. For the soft parts we write the product of two correlators defined as in \eqref{eq:correlator} and \eqref{eq:corrA}.

For the hard part we see on-shell propagation of the quark before the photon production and must thus insert a quark propagator before the first vertex. Considering the diagram in Fig.~\ref{fig:DY-twist3}, we may write:
\begin{equation}
  \frac{i \left( \bar{x} \slashed{P}_2 + (y-x) \slashed{P}_1 \right)}
       {(\bar{x} P_2 + (y-x) P_1)^2 + i\epsilon}.
\end{equation}
The relevant contribution to the denominator here is $2\bar{x}(y-x)P_1^+P_2^-=\bar{x}(y-x)s$. In the numerator we need only take the `+' component since the gluon comes from the proton and, using the fact that the LAB and CM frames correspond, $(P_1^+)^2=(P_2^-)^2$ and therefore $P_1^+=\sqrt{\frac{s}{2}}\,\hat{p}^+$. The propagator term then becomes
\begin{equation}
  i \frac{\slashed{\hat{p}}^+}{\bar{x}\sqrt{2s}}
  \frac{(y-x)}{(y-x) + i\epsilon}.
\end{equation}
Similarly, for the Hermitian conjugate of the diagram in Fig.~\ref{fig:DY-twist3} the propagator becomes the complex conjugate and the role of $x$ and $y$ are interchanged. We thus have:
\begin{equation}
  -i \frac{\slashed{\hat{p}}^+}{\bar{x}\sqrt{2s}}
  \frac{(x-y)}{(x-y) - i\epsilon}.
\end{equation}
Only these two terms contribute to the SSA owing to the pole in the denominator at $x=y$, or simply when the gluon carries zero momentum. This source of imaginary part can be regularised via Cauchy's theorem and used for the SSA calculation \cite{Efremov,Qiu}.

Let us now examine the complete expression for the hadronic part. We must evaluate the integral
\begin{multline}
  \int d^2\vec{q}_T \, W^{\mu\nu} =
  \frac{e^2}{N_c} \left\lbrace
  \Tr\!\bigg(
    \phi(x) \gamma^\mu \bar\phi(x) \gamma^\nu
  \vphantom{\frac00}\bigg)
  \vphantom{\frac00}\right.
\\
  + \int dy \, \left[ \Tr\!\bigg(
      \phi_A^\alpha (x,y) \gamma^\mu
      \frac{\slashed{\hat{p}}^+}{\bar{x}\sqrt{2s}}
      \frac{(y-x)}{(y-x)+i\epsilon} \gamma_\alpha \bar\phi(\bar{x})
      \gamma^\nu
    \bigg)
    \right.
\\
  \left.
  \left.
    \null + \Tr\!\bigg(
    \phi_A^\alpha (y,x) \gamma^\mu \bar\phi(\bar{x}) \gamma_\alpha \frac{\slashed{\hat{p}}^+}{\bar{x}\sqrt{2s}} \frac{(x-y)}{(x-y) + i\epsilon} \gamma^\nu
  \bigg)
  \right]
  \right\rbrace.
\label{eq:adronicopol}
\end{multline}
The first term comes from the diagram with no gluon exchange and contributes to the denominator of the SSA. In order to evaluate it we make the expansion
\begin{equation}
  \phi(x) =
  \tfrac{1}{2} \Big[
    V_\mu\gamma^\mu + A_\mu \gamma^5 \gamma^\mu
    + i T_{\mu\nu} \gamma^5 \sigma^{\mu\nu}
  \Big],
\end{equation}
where
\begin{subequations}
\begin{align}
  \tfrac{1}{2} V_\mu &= \tfrac{1}{2} \Tr\!\big( \gamma_\mu \phi \big)
  = f(x) P_\mu,
\\
  \tfrac{1}{2} A_\mu &= \tfrac{1}{2} \Tr\!\big( \gamma_5 \gamma_\mu \phi \big)
  = g_1(x) \lambda P_\mu + g_T(x) M S_{T\mu},
\\
  \tfrac{1}{2} T_{\mu\nu} &=
  \tfrac{1}{2} \Tr\!\big( i\sigma_{\mu\nu} \gamma_5 \phi \big)
\notag
\\
  &
  = - h_1(x) S_{T[\mu}P_{\nu]} + h_L(x) M \lambda P_{[\mu}n_{\nu]}.
\end{align}
\end{subequations}
Recalling that $\slashed{P}_1=P_1^+\gamma^-$\!, $\slashed{P}_2=P_2^-\gamma^+$\!, $\slashed{\hat{p}}=\gamma^-$ and $P_1^+P_2^-=s/2$, we finally obtain:
\begin{multline}
  \Tr\!\big( \phi(x) \gamma^\mu \phi(\bar{x}) \gamma^\nu \big)
  = 2s \Big[ f(x)f(\bar{x}) g_T^{\mu\nu}
\\
  + g_1(x) f(\bar{x}) \lambda i\epsilon^{-\mu+\nu}
  + g_T(x) f(\bar{x}) i\epsilon^{\sigma\mu+\nu} S_{1T\sigma}
  \Big].
\end{multline}
Note that the only the first term is symmetric and thus alone survives contraction with the leptonic tensor.

We now evaluate the second term of \eqref{eq:adronicopol}, containing the \mbox{twist-3} correlator. For a polarised hadron the full expansion for the correlator becomes
\begin{multline}
  \phi_A^\alpha(x,y) =
  \Big[
    i\epsilon^{\alpha-\mu+} S_{1\mu} \, G_A(x,y)
    + S_1^\alpha \, \tilde{G}_A(x,y) \, \gamma^5
\\
    + \Big(
      \lambda_1 H_A(x,y) \, \gamma^5 \gamma^\alpha +
      2E_A(x,y) \, \gamma^\alpha
    \Big)
  \Big] \slashed{P}_1
\label{eq:phiA}
\end{multline}
(recall that $\alpha$ is a transverse index), while for the unpolarised hadron we only have the vector term $f(\bar{x})\slashed{P}_1$. Using again $\slashed{P}_1=P_1^+\gamma^-$, $\slashed{P}_2=P_2^-\gamma^+$ and $\slashed{\hat{p}}^+=\gamma^-$, we then have
\begin{multline}
  \frac{1}{\bar{x} \sqrt{2s}} \int dy \, \frac{y-x}{y-x+i\epsilon} \,
  f(\bar{x})
\\
  \times
  \Big[
    G_A(x,y) \, i \epsilon_T^{\alpha\beta} S_{1\beta} P_1^+ P_2^-
    \Tr\!\big(
      \gamma^- \gamma^\mu \gamma^- \gamma_\alpha \gamma^+ \gamma^\nu
    \big)
\\
    + \tilde{G}_A(x,y) S_1^\alpha P_1^+ P_2^-
    \Tr\!\big(
      \gamma^5\gamma^-\gamma^\mu\gamma^-\gamma_\alpha\gamma^+\gamma^\nu
    \big)
  \Big].
\end{multline}
Evaluating traces and collecting the Levi--Civita tensor, we obtain the final expression
\begin{multline}
  - \frac{4i}{\bar{x}} \sqrt{\frac{s}{2}}
  \int dy \, \frac{y-x}{y-x+i\epsilon}
\\
  \times
  f(\bar{x}) \, \Big( G_A(x,y)- \tilde{G}_A(x,y) \Big) \,
  g^{-\mu} \epsilon_T^{\nu\alpha} S\IPhantHi_{1T\alpha}.
\end{multline}

Finally, we evaluate the third and last term of \eqref{eq:adronicopol}. Following the same procedure and summing this with other results, we obtain the complete expression:
\begin{multline}
  - \frac{4i}{\bar{x}} \sqrt{\frac{s}{2}}
  \bigg\lbrace
  \int dy \, g^{-\mu} \epsilon_T^{\nu\alpha} S\IPhantHi_{1T\alpha}
\\
  \hspace*{5em}
  \times
  \bigg[
    \frac{y-x}{y-x+i\epsilon}  \Big(G_A(x,y) - \tilde{G}_A(x,y)\Big)
\\
  \hspace*{7em}
    -\frac{y-x}{y-x-i\epsilon} \Big(G_A(y,x) - \tilde{G}_A(y,x)\Big)
  \bigg]
\\
  \hspace*{1em}
  + \int dy \, \frac{y-x}{y-x-i\epsilon} S\IPhantHi_{1T\alpha}
\\
  \times
  \Big(
    G_A(y,x) g\IPhantLo^{-\lbrace\mu} \epsilon_T^{\nu\rbrace\alpha}
    - \tilde{G}_A(y,x) \epsilon^{-\mu\alpha\nu}
  \Big)
  \bigg\rbrace.
\end{multline}
With our choice of gauge and using \eqref{eq:AtoF}, this is equivalent to the gauge-invariant expression:
\begin{multline}
  - \frac{4}{\bar{x}} \sqrt{\frac{s}{2}} \, f(\bar{x})
  \bigg\lbrace
    \int dy \, g^{-\mu} \epsilon_T^{\nu\alpha} S\IPhantHi_{1T\alpha}
\\
  \hspace*{5em}
    \times
    \bigg[
      \frac{-1}{y-x+i\epsilon} \Big(G_F(x,y) - \tilde{G}_F(x,y)\Big)
\\
  \hspace*{7em}
      -\frac{1}{y-x-i\epsilon} \Big(G_F(y,x) - \tilde{G}_F(y,x)\Big)
    \bigg]
\\
  \hspace*{1em}
    + \int dy \, \frac{1}{y-x-i\epsilon} S\IPhantHi_{1T\alpha}
\\
    \times
    \Big(
      G_F(y,x) g\IPhantLo^{-\lbrace\mu} \epsilon_T^{\nu\rbrace\alpha}
      -\tilde{G}_F(y,x) \epsilon^{-\mu\alpha\nu}
    \Big)
  \bigg\rbrace,
\end{multline}
where only the first two terms are symmetric and thus survive contraction with the leptonic part. We now regularise the pole via Cauchy's theorem and take the imaginary part to obtain the following expression:
\begin{multline}
  \frac{4}{\bar{x}} \sqrt{\frac{s}{2}} \, f(\bar{x}) \,
  \bigg\lbrace
    \int dy \, g^{-\mu} \epsilon_T^{\nu\alpha} S\IPhantHi_{1T\alpha}
\\
  \hspace*{1em}
    \times
      (-i\pi) \delta(y-x) \Big[ G_F(x,y) - \tilde{G}_F(x,y)
\\
  \hspace*{9.5em}
      - G_F (y,x) + \tilde{G}_F(y,x) \Big]
\\
    - \int dy \, i\pi \delta(y-x) \, G_F(y,x) \,
    g\IPhantLo^{-\lbrace\mu} \epsilon_T^{\nu\rbrace\alpha}
    S\IPhantHi_{1T\alpha}
  \bigg\rbrace.
\end{multline}
Note that the $\delta$-function sets the term inside square brackets to zero. Adding the first term of \eqref{eq:adronicopol}, performing the sum over colours and reinstating the factor $e^2/N_c$, we finally obtain:
\begin{multline}
  W^{\mu\nu} = \frac{e^2}{N_c} 4s f(\bar{x})
  \bigg[
    -\tfrac{1}{2} f(x) \, g_T^{\mu\nu}
\\
    -\frac{1}{\bar{x}\sqrt{2s}} \, G_F(x,x) \,
    g\IPhantLo^{-\lbrace\mu} \epsilon_T^{\nu\rbrace\alpha}
    S\IPhantHi_{1T\alpha}
  \bigg].
  \label{eq:Wmunu}
\end{multline}

The final operation to perform is contraction with the leptonic tensor, following formula \eqref{eq:ex}. We start by evaluating the first term in the collinear expansion \eqref{eq:ex}, \emph{i.e.} with $\vec{p}_T$ identically zero:
\begin{align}
  L_{\mu\nu} W^{\mu\nu} |_{\vec{p}_T=0}
  &= - \frac{e^2}{N_c} \, f(\bar{x})
  \bigg[
    2s f(x) \, L_{\mu\nu} \, g_T^{\mu\nu}
\notag
\\
  & \hspace*{1.3em}
    - \frac{4s}{\bar{x} \sqrt{2s}} \, G_F(x,x) \, L_{\mu\nu} \,
    g\IPhantLo^{-\lbrace\mu} \epsilon_T^{\nu\rbrace\alpha}
    S\IPhantHi_{1T\alpha}
  \bigg]
\notag
\\
  &=
  s \, \frac{e^2}{N_c} \, Q^2 f(\bar{x})
  \bigg[
    (1+\cos^2\theta) \, f(x)
\notag
\\
  & \hspace*{1.3em}
    - \frac{|S_{1T}|}{\bar{x} \sqrt{s}} \, \sin2\theta \sin\phi_1 \, G_F(x,x)
  \bigg].
\label{eq:Contraction}
\end{align}
Inserting now Eq.~\eqref{eq:Wmunu} into the second term of formula \eqref{eq:ex}, recalling that the derivative term vanishes and noting that only the $Z_{\{\mu}X_{\nu\}}$ and $Z_{\{\mu}Y_{\nu\}}$ terms in $L_{\mu\nu}$ survive, we find that this contribution is identical to the second term of Eq.~\eqref{eq:Contraction} and thus implies a factor two. With this final observation, to first order in the collinear expansion, we have:
\begin{multline}
  L_{\mu\nu} W^{\mu\nu} = s \, \frac{e^2}{N_c} \, Q^2 f(\bar{x})
  \bigg[
    (1 + \cos^2\theta) \, f(x)
\notag
\\
    -\frac{2|S_{1T}|}{\bar{x} \sqrt{s}} \, \sin2\theta \sin\phi_1 \, G_F(x,x)
  \bigg],
\end{multline}
which leads to the following SSA:
\begin{equation}
  A_N = - \, \frac{2}{Q}
  \frac{\sin2\theta \sin\phi_1}{1+\cos^2\theta}
  \frac{\sum_a e_a^2 f_a(\bar{x}) \, G_{F\,a}(x,x)}
       {\sum_a e_a^2 f_a(\bar{x}) \, f_a(x)}.
\end{equation}
\section{Conclusions}

As in \cite{Metz,Ma:2012ph}, we find it crucial in our calculation to perform the collinear expansion with due respect to all possible ensuing transverse-momentum dependence. However, the extra contribution generated is of the same sign and magnitude as the na\"{\i}ve part, thus leading to an overall factor of two with respect to the older calculations and four for the later. Unfortunately, the level of detail provided in \cite{Teryaev} renders full in-depth comparison impossible. We note too that, while an explanation is proposed in \cite{Metz}, the level of detail is again not sufficient to permit comparison. With regard to \cite{Ma:2012ph}, the origin of the extra piece there does not appear to be the same as that of \cite{Metz}. Indeed, the authors appear to claim that in earlier papers the error lies in incorrect expansion of the hadronic tensor whereas the authors of \cite{Metz} point to failure to expand the leptonic tensor.

In conclusion, we have repeated the calculation of the transverse SSA in DY, including the contribution arising from collinear expansion of the leptonic tensor, as suggested in \cite{Metz}, and find a factor of four with respect to the most recent calculations presented in \cite{Metz} and \cite{Ma:2012ph}; we thus find ourselves in agreement with the results of~\cite{Teryaev}. Unfortunately however, with the published detail available, we cannot point clearly to possible sources of error in the other papers.

\end{fmffile}

\end{document}